\documentclass[apj]{emulateapj}

\newcommand {\Lya}    {Ly$\alpha$}   
\newcommand {\Lyb}    {Ly$\beta$}    
\newcommand {\Lyg}    {Ly$\gamma$}   
\newcommand {\OVI}    {\ion{O}{6}}   
\newcommand {\NV}     {\ion{N}{5}}

\newcommand {\tnma}{\tablenotemark{a}}

\newcommand {\dndz}  {$d{\cal N}/dz$}
\newcommand {\etal}  {et~al.}

\shorttitle{S5\,0716$+$714: Fast Flare and Redshift Constraints}
\shortauthors{Danforth \etal}

\begin{document}

\title{A Fast Flare and Direct Redshift Constraint in Far-Ultraviolet Spectra of the Blazar S5\,0716$+$714
\footnote{Based on observations made with the NASA/ESA {\it Hubble Space Telescope}, obtained from the data archive at the Space Telescope Science Institute. STScI is operated by the Association of Universities for Research in Astronomy, Inc. under NASA contract NAS 5-26555.}}

\author{Charles W. Danforth, Krzysztof Nalewajko, Kevin France, \& Brian A. Keeney}

\affil{CASA, Department of Astrophysical \& Planetary Sciences, University of Colorado, 389-UCB, Boulder, CO 80309; danforth@casa.colorado.edu}


\begin{abstract}

The BL\,Lacertae object S5\,0716$+$714 is one of the most studied blazars on the sky due to its active variability and brightness in many bands, including VHE gamma rays.  We present here two serendipitous results from recent far-ultraviolet spectroscopic observations by the Cosmic Origins Spectrograph onboard the {\it Hubble Space Telescope}.  First, during the course of our 7.3 hour HST observations, the blazar increased in flux rapidly by $\sim40$\% ($-0.45\;{\rm mag\,h^{-1}}$) followed by a slower decline ($+0.36\;{\rm mag\;h^{-1}}$) to previous far-UV flux levels.  We model this flare using asymetric flare templates and constrain the physical size and energetics of the emitting region.  Furthermore, the spectral index of the object softens considerably during the course of the flare from $\alpha_{\nu}\approx-1.0$ to $\alpha_{\nu}\approx-1.4$.  Second, we constrain the source redshift directly using the $\sim30$ intervening absorption systems.  A system at $z=0.2315$ is detected in \Lya, \Lyb, \OVI, and \NV\ and defines the lower bound on the source redshift.  No absorbers are seen in the remaining spectral coverage ($0.2315<z_{Ly\alpha}\la0.47$) and we set a statistical upper bound of $z<0.322$ (95\% confidence) on the blazar.  This is the first direct redshift limit for this object and is consistent with literature estimates of $z=0.31\pm0.08$ based on the detection of a host galaxy.

\end{abstract}

\keywords{BL Lacertae objects: individual: S5\,0716$+$714, radiation mechanisms: nonthermal, galaxies: jets, galaxies: active, intergalactic medium, quasars: absorption lines, ultraviolet: general}

\section{Introduction}

Blazars are a class of AGN in which non-thermal continuum emission is relativistically beamed into our line of sight, rendering them extremely bright across many wavebands at cosmological distances.  S5\,0716$+$714 is one of the most extensively studied blazars in the sky.  It belongs to the BL\,Lacertae Object (BL\,Lac) subclass of blazars, which are known for their smooth non-thermal continua without clear emission lines.  This lack of identifying spectral features makes the source redshifts, and hence the physical characteristics of these astrophysically interesting objects, very challenging to determine unambiguously.

Spectroscopically, S5\,0716$+$714 was first studied in the optical band by \citet{Biermann81} who found a completely featureless spectrum.  No emission or absorption lines were ever identified in this source, despite several attempts \citep{Stickel93,VermeulenTaylor95,RectorStocke01,Finke08}.  \citet{Nilsson08} detected an excess in the point spread function of S5\,0716$+$714 which is consistent with a host galaxy at $z=0.31\pm0.08$.  This range is consistent with the \citet{Stickel93} identification of a pair of galaxies near the sight line at $z\simeq 0.26$, suggesting that the host is part of a galaxy group.  Detection of S5\,0716$+$714 in Very-High-Energy (VHE) gamma-rays by the MAGIC telescope led to different upper limits on the source redshift: $z\lesssim0.5$ \citep{Anderhub09}, and $z<0.21\pm 0.09$ \citep{Prandini10}, though these depend on the poorly-characterized cosmic infrared background.

Blazars are also known for variability.  Even amongst blazars, S5\,0716$+$714 is well known for its strong and persistent optical variability with a duty cycle of $\sim90\%$ \citep{Ostorero06}.  Intraday variability (IDV) is routinely measured in this source with the reported maximum variability rates of $\sim0.1-0.16$ magnitudes per hour \citep{Wagner96,Wu05,Villata00,Montagni06,Ostorero06,Nesci02} to as high as $0.38$ magnitudes per hour \citep{Chandra11}.  An extensive statistical analysis by \citet{Montagni06} showed that variability rates of order $\sim 0.1\,{\rm mag\,hr^{-1}}$ usually last for no more than two hours, and their duty cycle is only a few percent.  Asymmetric flares, with the increase rate on average faster than the decline rate, were found by \citet{Wagner96} and \citet{Nesci02}, see, however, \citet{Ghisellini97} and \citet{Montagni06}.  Quasi-periodic oscillations were claimed over short time intervals with various periods: $\sim 1$ day and $\sim 7$ days \citep{Quirrenbach91,Wagner96}, $\sim4$ days \citep{HeidtWagner96}, and even as fast as $\sim 15$ minutes \citep{Rani10}.  

\begin{figure*}[t]
  \epsscale{.8}\plotone{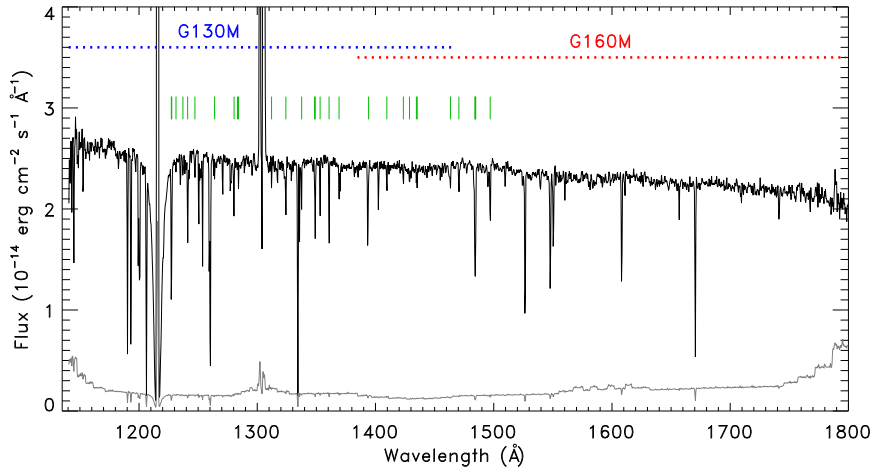}
  \caption{Coadded COS spectrum of S5\,0716$+$714.  Flux has been
  scaled to the first exposure (however, see discussion in
  Section~3.1) and smoothed by 35 pixels (approximately five
  resolution elements).  Shaded data shows coadded photon noise.
  Sharp absorption features are narrow absorbers; the locations of the
  30 intervening \Lya\ absorbers detected at $\ge5\sigma$ significance
  are indicated by green ticks.  Sharp emission features are
  geocoronal airglow.  The spectral ranges of the two
  medium-resolution, far-UV gratings on COS are shown.}
  \label{fig:overview}
\end{figure*}

BL\,Lac objects are extremely useful as probes of intervening material.  High signal-to-noise (S/N) optical and ultraviolet spectral observations are possible if the blazar is observed during a bright state.  Their smooth, power-law continua make it possible to detect weak, broad absorption features.  It was for this reason that S5\,0716$+$714 was observed by the Cosmic Origin Spectrograph on the {\it Hubble Space Telescope} (HST/COS; \citealt{Green12}).  The far-ultraviolet (FUV) spectrum of S5\,0716$+$714 provides an excellent dataset relevant to IGM cosmology which will be presented as part of a much larger study of the low-redshift IGM (Danforth \etal\ 2013, in prep).  

We present here two serendipitous discoveries from the S5\,0716$+$714 observations.  First, our five-orbit HST observations happened to observe a fast, asymmetric flare during which the blazar brightened by $>35\%$ and then faded to approximately the flux level at the start of our observations.  We discuss the flare and changes to the spectral energy distribution (SED) in Section~3.  Second, in Section~4, we use the observed intervening absorption systems (the \Lya\ forest) along the sight line to place the first direct constraints on the source redshift of S5\,0716$+$714.  We discuss and summarize our results in Section~5.

\section{Observations and Timing Analysis}

S5\,0716$+$714 was observed with the Cosmic Origins Spectrograph (COS) on December 27, 2011, as part of HST program 12025 (PI: Green).  Two exposures were made during each of five HST orbits over the course of 7.3 hours; the first five exposures with the G130M ($1135<\lambda<1450$ \AA, 6.00 ksec) grating, the second five with the G160M ($1400<\lambda<1795$ \AA, 8.25 ksec) grating.  The ten exposures were obtained from the Mikulski Archive for Space Telescopes (MAST) and reprocessed with a recent version of CALCOS (2.17.3a).  The calibrated, one-dimensional spectra were next coadded with the custom IDL procedures described in detail by \citet{Danforth10}.  We present the coadded COS spectrum of S5\,0716$+$714 in Figure~\ref{fig:overview} (however, see the cautions regarding coaddition of a strongly varying source in Section~3.1.) 

A Voigt profile fit to the Galactic \Lya\ profile in the G130M data gives a column density of $\log\,N(HI)=20.372\pm0.003$.  This corresponds to a color excess of $E(B-V)=0.045$ and we deredden all observed fluxes via \citet{Fitzpatrick99}\footnote{Note that the absorption-derived color excess measured here is $\sim50$\% larger than the extinction value derived from the low-resolution \citet{Schlegel98} dust maps.}.  

\begin{figure*}[t]
  \epsscale{.8}\plotone{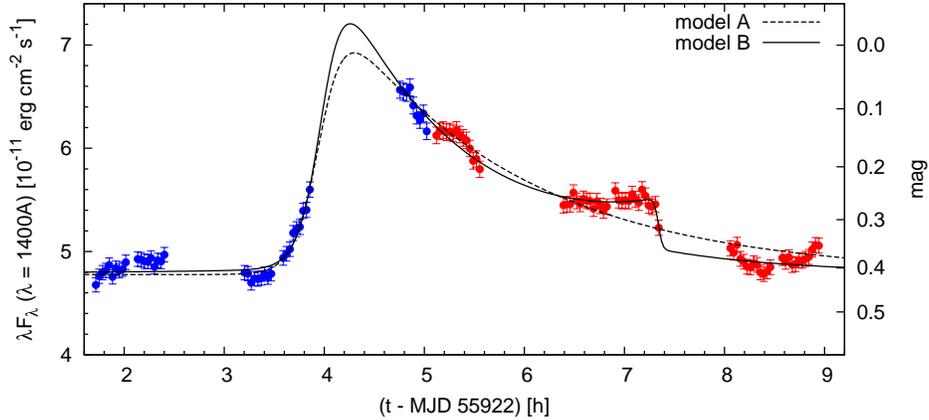} 
  \caption{Far-UV light curve of S5\,0716$+$714, presented as
  dereddened $\lambda F_\lambda$ values at $\lambda = 1400{\rm\AA}$
  extracted in 120~s time bins.  Blue and red points show data
  collected using the G130M and G160M gratings, respectively.  The
  overall light curve was fitted with two models consisting of one and
  two flare templates.  The right axis shows magnitude calculated
  relative to the flux density level of $7\times 10^{-11}\;{\rm
  erg\,s^{-1}\,cm^{-2}}$.}  \label{fig:lightcurve}
\end{figure*}

Many AGN show some degree of variability between one epoch and another and our coaddition software automatically scales exposures from different epochs to correct this (usually minor) effect.  However, during coaddition of the S5\,0716$+$714 data, it was noticed that the mean flux varied by as much as 35\% during the course of 7.3 hours.  COS is a photon-counting instrument; the arrival time of each photon is recorded as well as the two-dimensional position on the detector.  It is therefore possible to extract very high resolution ($\sim30$ ms) time-domain data \citep{France10}.  We extract an [$x_i$,$y_i$,$t_i$] photon list from each exposure $i$ and coadd these to create a master [$x$,$y$,$t$] photon list.  The total number of counts in a [$\Delta$$x$,$\Delta$$y$] box is integrated over a timestep $\Delta t$.  We use a timestep of $\Delta t=120$~s for S5\,0716$+$714 to balance time-resolution with reasonable photometric precision.  Continuum count rates are extracted over $\sim30$ (37)~\AA\ integration regions for the G130M (G160M) grating at four nominal far-UV wavelengths ($1190=[1178.6-1208.5$ \AA] and $1390=[1377.8-1407.7$ \AA] in G130M; $1460=[1447.9-1484.6$ \AA] and $1520=[1507.3-1544.0$ \AA] in G160M).  Narrow interstellar and intergalactic absorption lines are present in the data (see Figure~\ref{fig:overview}), however these features only remove a small number of photons relative to the bright continuum emission on 30~\AA\ scales, therefore ignoring narrow absorbers does not significantly impact the photometry produced here.  The instrumental background is computed in a similar manner, with extraction boxes offset below the active science region of the detector.  The instrumental background contributes $<0.6\%$ of the total photons measured in each extraction region. 

So that photometry from the blue G130M and red G160M gratings can be compared, flux-callibrated spectra for each exposure ($\sim20-28$ min) are fitted with a power law of the form $f(\lambda)=f(\lambda_0)\times(\lambda/\lambda_0)^{\alpha_\lambda}$ normalized at $\lambda_0=1400$ \AA, a wavelength covered by both gratings.  Raw light curves in units of counts $\rm s^{-1}$ are extracted in the wavelengths intervals defined above and in 2-min time intervals.  Flux calibrated light curves at $\lambda_0$ are obtained by dividing the raw light curves by exposure-averaged count rates in the same wavelengths intervals, and by multiplying them by $f(\lambda_0)$. The result is a nearly-continuous, flux-calibrated light curve for S5\,0716$+$714 (Figure~\ref{fig:lightcurve}) at 1400 \AA, interupted only by four periods of Earth occultation.  Henceforth, all times are quoted in hours starting from MJD\,55922 (0:00\,UT on December 27, 2011).  

\section{A Fast Flare in the Far-Ultraviolet}

Figure~\ref{fig:lightcurve} shows the strong source variability during our five-orbit ($7.3\;{\rm h}$) HST observing period; the blazar brightened by a factor of $\sim40\%$ ($0.36\;{\rm mag}$) before fading to very nearly the initial intensity.  Most dramatically, the source showed a monotonic flux increase of $0.17\;{\rm mag}$ during $0.39\;{\rm h}$, with the average variability rate of $-0.45\;{\rm mag\,h^{-1}}$ during the second orbit.  During the third orbit, we observed a quasi-monotonic flux decrease of $+0.11\;{\rm mag}$ over $0.26\;{\rm h}$, with an average variability rate of $+0.36\;{\rm mag\;h^{-1}}$.

Even though there are gaps in our light curve due to Earth occultation, the variability appears to be fairly simple.  Hence, we model the light curve with a constant background flux level $F_0$ and a single flare template which is described by a function
\begin{equation}
f_i(t) = \frac{2\,F_i}{\exp\bigl(\frac{t_i-t}{T_{\rm r,i}}\bigr) + \exp\bigl(\frac{t-t_i}{T_{\rm d,i}}\bigr)}
\end{equation}
with four parameters: flare normalization $F_i$, flux raising time scale $T_{\rm r,i}$, flux decay time scale $T_{\rm d,i}$, and epoch $t_i$ corresponding roughly to the flare maximum \citep{Abdo10}.  This Model~A is shown in Figure~\ref{fig:lightcurve}, and its parameters are reported in Table~1. The fit quality is mediocre ($\bar{\chi}^2=5.1)$ and the largest residuals are observed during the late stage of the flare.  Hence, we consider a more detailed Model~B, which includes a second flare template.  This model provides a substantial improvement in the fit quality ($\bar{\chi}^2=3.2)$, and the remaining residuals do not show any long-term structure.  The main difference in the main flare parameters in Model~B, compared to Model~A, are higher flare amplitude $F_1$ and shorter flux decay time scale $T_{\rm d,1}$.  The background flux level $F_0$, flux raising time scale $T_{\rm r,1}$, and the moment of peak flux $t_1$ are consistent between models A and B.  The second flare template $f_2(t)$ of Model~B has poorly constrained characteristic time scales; in particular, the very fast decay time $T_{\rm d,2}$ is almost entirely determined by the final 120~s bin at the end of the fourth orbit.  In Section 5, we discuss the physical implications of the main flare parameters.

\begin{deluxetable}{ccc}
\tabletypesize{\footnotesize}
\tablecolumns{3} 
\tablewidth{0pt} 
\tablecaption{Flare Model Parameters\tnma}
\tablehead{\colhead{Parameter}   &
           \colhead{Model A}    &
	   \colhead{Model B}    }
	\startdata
$F_0     $&$ 4.78\pm0.03 $&$ 4.79\pm0.02$\\
$F_1     $&$ 1.37\pm0.07 $&$ 1.59\pm0.12$\\
$t_1     $&$ 3.97\pm0.05 $&$ 3.99\pm0.05$\\
$T_{r,1} $&$ 0.13\pm0.04 $&$ 0.12\pm0.03$\\
$T_{d,1} $&$ 1.85\pm0.12 $&$ 1.28\pm0.12$\\
$F_2     $&    \nodata    &$ 0.25\pm0.04$\\ 
$t_2     $&    \nodata    &$ 7.34\pm0.02$\\
$T_{r,2} $&    \nodata    &$ 1.40\pm0.53$\\
$T_{d,2} $&    \nodata    &$ 0.02\pm0.02$\\
$\bar{\chi}^2$ & 5.1      &  3.2 \\
	\enddata
\tablenotetext{a}{Flux density in $10^{-11}\rm~erg~cm^{-2}~s^{-1}$ and times in hours.  Peak times are in hours since MJD\,55922.}
\end{deluxetable}

\subsection{Spectral Energy Distribution}

\begin{figure}
  \epsscale{1}\plotone{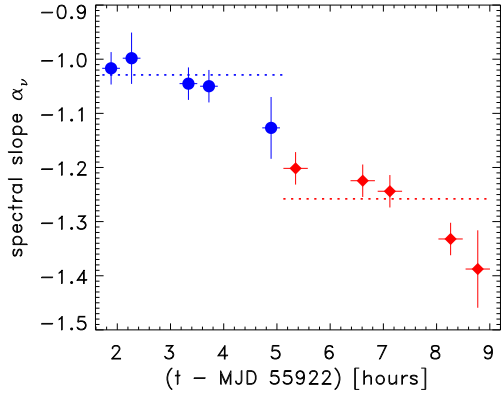}
  \caption{Power law fits to the individual COS observations show a
  softening of the SED (decreasing $\alpha_\nu$) throughout the HST
  observation before, during and after the flare.  Blue circles show
  fits to the G130M ($1140<\lambda<1450$ \AA) exposures while red
  diamonds denote G160M ($1400<\lambda<1795$ \AA) exposures.  Dotted
  blue and red lines show the fitted index for coadded G130M and G160M
  exposures.}
  \label{fig:alpha}
\end{figure}

Broad spectral coverage allows us to monitor any changes in the spectral index of S5\,0716$+$714.  Though the spectral ``throw'' of any individual observation is relatively small ($\Delta \lambda/\lambda\sim23\%$), they show significantly different spectral indices during the course of the HST observations.  We convert our measured $\alpha_\lambda$ values (Section~2) to the more conventional $\alpha_\nu$ via the relationship $\alpha_\nu=-(2+\alpha_\lambda)$ \citep{Shull12}.  The SED softens monatonically at the rate of $\Delta\alpha_{\nu}\approx-0.05\;{\rm h^{-1}}$ from $\alpha_\nu\approx-1.0$ at the beginning of the observation to $\alpha_\nu\approx-1.2$ near the flare maximum to $\alpha_\nu\approx-1.4$ at the end of the observation as shown in Fig.~\ref{fig:alpha}.  This trend is consistent with measured flux ratios from the blue and red ends of each spectral range.  


The highest-quality spectra can often be obtained by scaling the individual exposures and coadding them onto a common wavelength scale.  However, the significant change in the spectral index of S5\,0716$+$714 during the observations makes any SED derived from the combined observations suspect.  Combining just the observations in the same far-UV waveband produce slopes of $\alpha_\nu\approx-1.03$ and $-1.26$ for G130M and G160M channel coadditions, respectively (dotted horizontal lines in Figure~\ref{fig:alpha}).  Both slope values are typical of those of the individual observations, but we caution against trying to determine an overall SED from the combined far-UV observations.

\section{Redshift of S5\,0716$+$714}

Determining the source redshift of featureless AGN can be difficult, especially at higher redshift where the host galaxy cannot be easily identified.  Redshift limits can be inferred through indirect means (host galaxy non-detection, gamma ray emission, etc.).  A direct lower limit to the source redshift can be determined through intermediate absorption lines in moderate-resolution data (usually in the UV).  An automated line-finding algorithm (Danforth \etal\ 2013, in prep) finds $\sim30$ intervening \Lya\ absorption systems detected to greater than $5\sigma$ significance in the COS spectrum of S5\,0716$+$714.  Figures~1 and 4 show the observed COS spectral range with detected \Lya\ systems along the top edge in both wavelength and redshift space.  The reddest of these lines is at 1497 \AA, corresponding to $z_{\rm abs}=0.2315$, and is confirmed by corresponding detections in higher-order Lyman lines (\Lyb, \Lyg) as well as metal ions (\OVI, \NV) thus setting a firm lower limit on the redshift of S5\,0716$+$714.  The remaining spectral range ($\rm 1497\AA < \lambda < 1795\AA$; $0.232 < z_{Ly\alpha} < 0.47$) is devoid of \Lya\ absorption to a $5\sigma$ limiting equivalent width of $\sim 38$ m\AA\ ($\log\,N_{\rm HI}\sim 12.8$).  This technique was first used in \citet{Danforth10} to constrain the upper redshift limit of 1ES\,1553$+$113 and we refine the methodology here.

\begin{figure}[b]
  \epsscale{1.2}\plotone{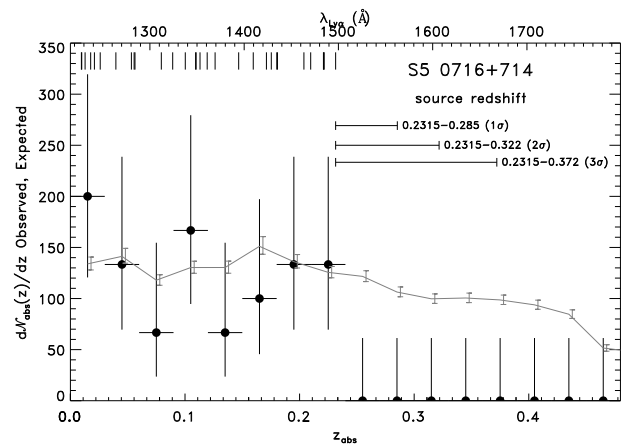}
  \caption{Intervening absorption lines provide the source redshift of
  S5\,0716$+$714.  Vertical ticks at the top of the figure show the
  wavelength/redshift location of $\sim30$ intervening \Lya\ systems
  detected at $\ge5\sigma$ significance in absorption.  The line
  density as a function of redshift, $d{\cal N}_{\rm abs}(z)/dz$ is
  shown in bins of $dz=0.03$.  The expected line density based on the
  quality of the COS data and the distribution of absorbers at
  low-redshift in a statistical sample \citep{DS08,Tilton12} is shown
  in gray.  We truncate the statistical distribution at different
  redshifts and compare to the observed distribution with a K-S test.
  The upper limit is constrained to $z<0.322$ ($z<0.372$) to a
  $2\sigma$ ($3\sigma$ confidence limit).  The strong absorption
  system at $z=0.2315$ defines the lower redshift limit.}
  \label{fig:zrange}
\end{figure}

We calculate the expected density of intervening \Lya\ systems per unit redshift, $d{\cal N}(z)/dz$ (shaded curve in Figure~\ref{fig:zrange}).  The minimum column density absorber that can be detected at a $4\sigma$ level is predicted based on the S/N in the data \citep{Keeney12} and the \dndz\ at that limiting column is drawn from the large, low-redshift IGM surveys of \citet{DS08} and \citet{Tilton12}.  We assume no evolution in the \Lya\ forest over this small redshift, but including a modest evolution produces negligible changes in the predicted absorber distribution.  The overall density of absorbers toward S5\,0716$+$714 (black data points) follows the predicted density consistently though with variations in individual redshift bins most likely due to cosmic variance.  

To constrain the upper limit on source redshift, we truncate the predicted \dndz\ distribution at a range of redshifts and compare this modeled absorber distribution to the observed distribution using a K-S test.  S5\,0716$+$714 can be constrained to $0.2315<z\la 0.285$ (68\% confidence), $0.2315<z\la 0.322$ (95\%), or $0.2315<z\la 0.372$ (99.7\%).

Weak \Lya\ emission is occasionally seen from low-redshift BL\,Lac objects \citep{Stocke11} and is an unambiguous, direct way of determining the redshift of blazars to high precision.  The only deviations from a smooth power-law continuum seen in the coadded G160M data are likely artifacts produced by the coadding individual exposures with rapidly changing spectral index and none is convincing as \Lya\ emission at $z>0.2315$.  We set a $3\sigma$ upper limit of $I(\rm Ly\alpha)\la 5\times10^{-15}~erg~cm^{-2}~s^{-1}$ ($EW\la115$~m\AA) by measuring the equivalent width detection limit \citep{Keeney12} for an emission feature with FWHM$\approx3$~\AA\ in the rest frame (typical of those seen by Stocke \etal\ in Mrk\,421, PKS\,2005$-$489 and Mrk\,501).  This intensity limit translates to an isotropic luminosity of $L(\rm Ly\alpha)\la 6.5\times10^{41}~erg~s^{-1}$.  All three \Lya\ emission detections at low-$z$ in \citealt{Stocke11} are considerably below this level, so this is not a terribly constraining upper limit.  Furthermore, high-$z$ AGN (of any sort) have been observed with luminosities as low as $\rm 10^{41}~erg~s^{-1}$ \citep{DijkstraWyithe06}.  

\section{Discussion}

We obtained the most sensitive, highest-resolution far-UV spectroscopic observation of S5\,0716$+$714 to date.  The precise source redshift remains unknown.  However, the detection of narrow \Lya\ forest features enables the determination of a new, independent constraint --- $0.2315<z<0.322$ at a $2\sigma$ confidence level.  This redshift range is consistent with the result of \citet{Nilsson08}, $z=0.31\pm0.08$, based on the photometric detection of the host galaxy as well as the intriguing possibility that S5\,0716$+$714 is associated with a group of nearby galaxies observed at redshift $z=0.26$ \citep{Stickel93}.  We note that the uncertainty in our estimate is based on the statistical properties of the low-redshift \Lya\ forest along many sight lines, while the uncertainty of the \citet{Nilsson08} estimate is based on the statistics of the host galaxy luminosities for BL\,Lacs.  Our source redshift estimate is also consistent with the upper limit $z<0.21\pm0.09$ obtained by \citet{Prandini10} from modeling the source SED at VHE gamma-rays and also with the $z\lesssim0.5$ estimate of \citep{Anderhub09}. 

Our observations of S5\,0716$+$714 coincided with a period of rapid flux variability.  This is not surprising, since this source is well known for almost uninterrupted variability at multiple time scales.  We detected episodes of very fast variability rate, up to $0.45\;{\rm mag\,h^{-1}}$.  This is higher than the maximum observed variability rate found in the most extensive study of IDV in this source \citep{Montagni06}, but comparable to the variability rate found by \citet{Chandra11} during one night out of five.  It seems that such fast variability rates are limited to periods of a fraction of an hour.  This appears to be consistent with the general trend that faster variability rates last for shorter time intervals (see Figure~7 in \citealt{Montagni06}).  It is also known that the frequency of light curve segments with constant variability rate decreases exponentially with increasing variability rate.  This would indicate rapid flares such as the one we observed with HST/COS, while spectacular, are probably not important in the overall source energetics.  However, the study of \citet{Montagni06} uses much longer sampling rates, and to obtain a statistical picture of intra-hour variability of this blazar one needs a much larger sample of high-cadence observations.

The decomposition of the observed light curve of S5\,0716$+$714 into a coherent flaring component and a quasi-static background (Model~B, above) allows us to constrain the physics of the emitting region producing the flare.  The rise time of the flare, $T_{\rm r,1}=(7.4\pm 1.8)\;{\rm min}$, indicates the emitting region radius 
\begin{equation}
R \lesssim \frac{\mathcal{D}cT_{\rm r,1}}{(1+z)} \simeq 3.4\times 10^{-5}\;{\rm pc}\times\left(\frac{\mathcal{D}}{10}\right)\left(\frac{T_{\rm r,1}}{7.4\;{\rm min}}\right)
\end{equation}
(for $z=0.26$), where $\mathcal{D}$ is the Doppler factor of the emitting region.  The flux decay time constrains the cooling time scale 
\begin{equation}
T_{\rm d,1} \gtrsim t_{\rm cool,obs} = \frac{3m_{\rm e}c(1+z)}{4\sigma_{\rm T}\mathcal{D}u_0'\gamma_{\rm e}}\,,
\end{equation}
where $u_0'=u_B'+u_{\rm rad}'=(1+q)B'^2/(8\pi)$ is the co-moving energy density of the magnetic field (determining the synchrotron cooling rate) plus the diffuse radiation (determining the inverse-Compton cooling rate), $q=u_{\rm rad}'/u_{\rm B}'\simeq L_{\rm IC}/L_{\rm syn}$ is the Compton dominance parameter (which for BL Lacs is of order of unity), and $\gamma_{\rm e}$ is the characteristic random Lorentz factor of electrons producing the observed FUV emission.  

Since the FUV continuum emission is produced by the synchrotron process, we use the expression for the observed frequency of the synchrotron radiation, 
\begin{equation}\nu_{\rm syn,obs} = \frac{0.274\;\mathcal{D}e\gamma_e^2B'}{(1+z)m_{\rm e}c}\simeq\frac{c}{1400\;{\rm\AA}},\end{equation}
to eliminate $\gamma_{\rm e}$ and obtain an estimate of the magnetic field strength in the emitting region
\begin{eqnarray}
B' &\gtrsim& (0.274m_{\rm e}ce)^{1/3}\times\left(\frac{6\pi}{\sigma_{\rm T}}\right)^{2/3}\left[\frac{(1+z)}{\mathcal{D}\nu_{\rm syn,obs}}\right]^{1/3}\nonumber\\ &&\times[(1+q)T_{\rm d,1}]^{-2/3}
\nonumber\\
&\simeq& 2\;{\rm G}\times\left(\frac{\mathcal{D}}{10}\right)^{-1/3}(1+q)^{-2/3}\nonumber\\ &&\times\left(\frac{\nu_{\rm syn,obs}}{c/1400\;{\rm\AA}}\right)^{-1/3}\left(\frac{T_{\rm d,1}}{1.28\;{\rm h}}\right)^{-2/3}\,.
\end{eqnarray}
It follows that 
\begin{eqnarray}
\gamma_{\rm e} &\lesssim& \left(\frac{\sigma_{\rm T}m_{\rm e}c}{6\pi}\right)^{1/3}(0.274e)^{-2/3}\nonumber\\ &&\times\left[\frac{(1+z)(1+q)T_{\rm d,1}}{\mathcal{D}}\right]^{1/3}\nu_{\rm syn,obs}^{2/3}
\nonumber\\
&\simeq& 5300\left(\frac{\mathcal{D}}{10}\right)^{-1/3}(1+q)^{1/3}\nonumber\\ &&\times\left(\frac{\nu_{\rm syn,obs}}{c/1400\;{\rm\AA}}\right)^{2/3}\left(\frac{T_{\rm d,1}}{1.28\;{\rm h}}\right)^{1/3}\,.
\end{eqnarray}
The apparent flare luminosity is 
\begin{equation}L_1 = 4\pi d_{\rm L}^2F_1 \simeq 3.2\times 10^{45}\;{\rm erg\,s^{-1}}. \end{equation}
We use it to estimate the co-moving energy density of the synchrotron radiation 
\begin{eqnarray}
u_{\rm syn}' &=& \frac{L_1}{4\pi c\mathcal{D}^4R^2} \gtrsim \frac{1}{4\pi c^3}\times\frac{(1+z)^2L_1}{\mathcal{D}^6T_{\rm r,1}^2}
\nonumber\\
&\simeq& 78\;{\rm erg\,cm^{-3}}\left(\frac{\mathcal{D}}{10}\right)^{-6}\left(\frac{T_{\rm r,1}}{7.4\;{\rm min}}\right)^{-2}\nonumber\\ &&\times\left(\frac{L_1}{3.2\times 10^{45}\;{\rm erg\,s^{-1}}}\right)\,.
\end{eqnarray}
In a typical BL Lac, one can neglect the external radiation, and thus the total energy density of the diffuse radiation is $u_{\rm rad}'\simeq u_{\rm syn}'$.  Using the estimate for $B'$, we can write a direct relation between the Compton dominance parameter and the Doppler factor: 
\begin{equation}q/(1+q)^{4/3}\simeq 485\,(\mathcal{D}/10)^{-16/3}.\end{equation}
We can further use the apparent flare luminosity to estimate the number of electrons contributing to the FUV emission 
\begin{eqnarray}
N_{\rm e}&\simeq& 6\pi L_1/(\sigma_{\rm T}c\mathcal{D}^4B'^2\gamma_{\rm e}^2)\\ \nonumber
         &\simeq& 2.7\times 10^{48}\times (\mathcal{D}/10)^{-8/3}(1+q)^{2/3},
\end{eqnarray} 
and their co-moving energy density 
\begin{eqnarray}
u_{\rm e}'&\simeq& 3N_{\rm e}\gamma_{\rm e}m_{\rm e}c^2/(4\pi R^3)\\ \nonumber
          &\simeq& 2400\;{\rm erg\,cm^{-3}}\times (\mathcal{D}/10)^{-6}(1+q)\,.
\end{eqnarray}

The ratio of the co-moving electron to synchrotron energy density is thus 
\begin{eqnarray}
\frac{u_{\rm e}'}{u_{\rm syn}'} &\simeq& \frac{3(1+q)T_{\rm d,1}}{T_{\rm r,1}} \\ \nonumber
&\simeq& 31\,(1+q)\left(\frac{T_{\rm d,1}}{1.28\,{\rm h}}\right)\left(\frac{7.4\,{\rm min}}{T_{\rm r,1}}\right)\,.
\end{eqnarray}
The fact that $u_{\rm e}'\gg u_{\rm syn}'$ is the consequence of $T_{\rm d,1}\gg T_{\rm r,1}$.  Since we assumed that $T_{\rm d,1}\simeq t_{\rm cool,obs}$ (observed cooling time scale) and $T_{\rm r,1}\simeq t_{\rm lc,obs} = (1+z)R/(\mathcal{D}c)$ (observed light crossing time scale), the ratio of $(u_{\rm e}'/u_{\rm syn}')$ is in fact determined by the ratio of $(t_{\rm cool,obs}/t_{\rm lc,obs})$.

If we assume equipartition between electrons and the magnetic field, $u_{\rm e}'\simeq u_B'$ \citep{Bottcher09}, we will obtain a very large Doppler factor, $\mathcal{D}\simeq 62$, and a very small Compton dominance parameter $q\simeq 0.031$.  Such a scenario poses a severe problem of explaining the extremely efficient acceleration of the emitting region, but it predicts a negligible gamma-ray signature of the FUV flare via the Synchrotron Self-Compton (SSC) mechanism.  If we instead assume that $q=1$, we will still require a large Doppler factor, $\mathcal{D}\simeq 38$, but also a strong departure from the equipartition, with $u_{\rm e}'/u_{\rm B}'\simeq 62$.  It should be noted that gamma-ray emission was detected in S5\,0716$+$714 by MAGIC \citep{Anderhub09}, AGILE \citep{Vittorini09} and {\it Fermi} at the level of $L_\gamma\sim (0.5-5)\times 10^{46}\;{\rm erg\,s^{-1}}$.  Hence, the Compton dominance parameter $q$ cannot be much larger than unity and some extreme physical parameter of the emitting region producing the fast FUV flare is unavoidable.  This makes our case of a fast FUV flare similar to the fast gamma-ray flares observed in other BL\,Lacs (PKS\,2155$-$304, \citealt{Aharonian07}; Mrk\,501, \citealt{Albert07}) and one FSRQ (PKS\,1222$+$216, \citealt{Aleksic11}).  They also seem to require extremely high Doppler factors \citep{Begelman08}, and a number of theoretical scenarios were proposed to explain them \citep{Levinson07, SternPoutanen08, GhiselliniTavecchio08, Giannios09, LyutikovLister10, Nalewajko11}.  In the case of S5\,0716$+$714, we cannot constrain the distance of the emitting region from the central black hole, and thus we have greater freedom of theoretical scenarios than in the case of PKS\,1222$+$216 \citep{Tavecchio11,Nalewajko12}.

We observed strong spectral variability with the systematic spectral index change of $\Delta\alpha_\nu=-0.4$ over the course of $\sim 7\;{\rm h}$.  Strong intraday spectral variations in the optical/UV band is rarely reported. For example, \cite{Wu12} found the $B-R$ color to vary with the amplitude of $|\Delta(B-R)|\simeq 0.08$ within an hour. This corresponds to $|\Delta\alpha_\nu|=|\Delta(B-R)|/0.413\simeq 0.2$, which means a significantly faster variability rate.  However, we have found no case of a systematic spectral index variation over several hours.

\medskip
Spectroscopic observations of blazars are proving to be versatile tool with which to study a range of astrophysical processes from galactic structure to cosmology to the study of AGN processes.  Our observations of S5\,0716$+$714 illustrate the power of time-domain sensitivity as well.  Not only is the Cosmic Origins Spectrograph the most sensitive far-UV spectroscopic instrument ever flown, but the temporal resolution can be exploited to look for microvariability on very short time scales.  A number of other blazars have been observed with HST/COS as part of a larger program on low-redshift cosmology.  Many cover only a short time period (1-3 HST orbits), but those with similar 5-orbit observations do not show appreciable variability.  We will continue to both look for notable variability in AGN observations and to determine the source redshifts of poorly-constrained blazars through spectroscopy of their intervening absorbers.

\medskip\medskip

The authors wish to acknowledge very inciteful discussions with John Stocke during this analysis as well as several good suggestions from Eric Perlman and our anonymous referee.  The authors made extensive use of the MAST, ADS, and IGwAD Archives during this work.  CD, KF, and BK were supported by NASA grants NNX08AC146 and NAS5-98043 to the University of Colorado at Boulder.  KN was supported by the NSF grant AST-0907872, the NASA ATP grant NNX09AG02G, and the Polish NCN grant DEC-2011/01/B/ST9/04845.  

{\it Facility: HST (COS)}

\end{document}